# 1D Flat Bands in Phosphorene Nanoribbons with Pentagonal Nature

*Shuo Sun,[1,2,#] Jing-Yang You,[3,#] Zhihao Cai,[4,5,#] Jie Su,[2,#] Tong Yang,[6] Xinnan Peng,[2] Yihe Wang,[2,7] Daiyu Geng,[4,5] Jian Gou,[3] Yuli Huang,[7] Sisheng Duan,[3] Lan Chen,[4,5,8] Kehui Wu,[4,5,8,9] Andrew T. S. Wee,[3] Yuan Ping Feng,[3] Jia Lin Zhang,[10]* Jiong Lu,[2]* Baojie Feng,[4,5,9]* and Wei Chen[2,3,7]**

[1]Department of Physics, Shanghai Key Laboratory of High Temperature Superconductors, Institute for Quantum Science and Technology, Shanghai, 200444, China

[2]Department of Chemistry, National University of Singapore, 3 Science Drive 3, 117543, Singapore

[3]Department of Physics, National University of Singapore, 2 Science Drive 3, 117551, Singapore

[4]Institute of Physics, Chinese Academy of Sciences, Beijing 100190, China

[5]School of Physical Sciences, University of Chinese Academy of Sciences, Beijing, 100190, China

[6]Department of Applied Physics, The Hong Kong Polytechnic University, Hung Hom, Hong Kong SAR, China

[7]Joint School of National University of Singapore and Tianjin University, International Campus of Tianjin University, Binhai New City, Fuzhou 350207, China

[8]Songshan Lake Materials Laboratory, Dongguan, Guangdong, 523808, China

[9]Interdisciplinary Institute of Light-Element Quantum Materials and Research Center for Light-Element Advanced Materials, Peking University, Beijing, 100871, China

[10]School of Physics, Southeast University, Nanjing 211189, China

These authors contributed equally: S. S., J. Y., Z. C. and J. S.

*Corresponding Authors: Jia Lin Zhang, Jiong Lu, Baojie Feng, and Wei Chen

**Email:** J. L. Z., (phyzjl@seu.edu.cn), J. L., (chmluj@nus.edu.sg) B. F. (bjfeng@iphy.ac.cn) and W. C. (phycw@nus.edu.sg).

**Abstract:** Materials with flat bands can serve as a promising platform to investigate strongly interacting phenomena. However, experimental realization of ideal flat bands is mostly limited to artificial lattices or moiré systems. Here we report a general way to construct one-dimensional (1D) flat bands in phosphorene nanoribbons (PNRs) with pentagonal nature: penta-hexa-PNRs and penta-dodeca-PNRs, wherein the corresponding 1D flat bands are directly verified by using angle-resolved photoemission spectroscopy. We confirm that the observed 1D flat bands originate from the electronic 1D zigzag and Lieb lattices, respectively, as revealed by the combination of bond-resolved scanning tunneling microscopy, scanning tunneling spectroscopy, tight-binding models, and first-principles calculations. Our study demonstrates a general way to construct 1D flat bands in 1D solid materials system, which provides a robust platform to explore strongly interacting phases of matter.

## 1. Introduction

Flat bands,[1] being dispersionless in reciprocal space, feature macroscopically degenerate eigenstates and vanishing group velocity. This results in high sensitivity to any relevant perturbations, and even weak perturbations can lift the degeneracy and bring various strongly correlated phenomena and exotic topological phases, such as the high-temperature superconductivity,[2] fractional quantum hall effect,[3, 4] anomalous Landau level,[5, 6] and so on. The intriguing physics of flat bands has generated considerable excitements over the years and enormous experimental efforts have been devoted to verifying the existence of flat bands.[7-9] However, the experimental progresses are mainly limited to engineered materials, including moiré systems[8, 10-15] and artificial systems.[7, 16-20] Besides, the relatively large unit cell size in moiré systems leads to a low electron density, which hinders the type of physics associated with high electron density.[21] Therefore, it is vital to rationally construct flat bands in non-moiré solid materials, and further to explore the corresponding quantum states.[22-24]

Flat bands can be generally divided into two types: localized states and itinerant states with destructive interference. The latter is rare but exceedingly desired, where the wavefunctions are confined in a specific region, resulting in the compact localized states.[25-29] In general, stringent symmetry and coupling are required to induce destructive interference, making flat band systems so rare in solid world. Prototypical examples of constructing flat bands based on destructive interference include kagome and Lieb lattices, in which only nearest-neighbor hopping is taken into consideration. Although various kagome/Lieb nanostructures have been synthesized, their corresponding electronic flat bands have not been observed because of the multidimensional and complex electron hoppings.[30, 31] For the reported kagome materials, the corresponding kagome layers are usually embedded within the bulk phases,[32-41] wherein interlayer couplings are non-negligible, resulting in the disappearance of ideal flat bands in these systems. To this point, dimensionality reduction to 1D is a feasible way for controlling the hopping degree of freedom, which is in favor of giving rise to ideal flat bands. On the other hand, the realization of flat bands in 1D solid materials is highly desired due to the miniaturization of energy-efficient quantum devices, but remains lacking.

Here we report a direct on-surface synthesis and realization of 1D flat bands in two different phosphorene nanoribbons (PNRs) systems on Ag(110), penta-hexa-PNRs and penta-dodeca-PNRs, which can be approximated as electronic 1D zigzag and Lieb lattices, respectively. The specific symmetries of 1D PNRs lead to the formation of flat bands. The atomically resolved structures and the resulting flat bands have been precisely revealed through a combination of bond-resolved scanning tunneling microscopy (BR-STM), scanning tunneling spectroscopy (STS), angle-resolved photoemission spectroscopy (ARPES), as well as tight binding (TB) models and first-principles density functional theory (DFT) calculations.

## 2. Results and Discussion

### 2.1. 1D Zigzag and Lieb Lattices Models

Based on the general picture of constructing ideal flat bands, the key component is constraining the hoppings within 1D systems with better tunability, which requires specifically designed lattices with certain geometries. Figure 1A and 1B are the typical 1D zigzag lattice and 1D Lieb lattices, respectively. Due to the 1D nature and the special structure of zigzag and Lieb the lattices, the flat bands can be generated. As for the 1D zigzag lattice (Figure 1A), the effective Hamiltonian can be expressed as:

$$\boldsymbol{H} = \begin{pmatrix} 2t_2\cos(k) & 2t_1\cos(k/2) \\ 2t_1\cos(k/2) & 2t_2\cos(k) \end{pmatrix}.$$

In the tight-binding model of the zigzag structure, considering only the nearest-neighbor hopping—while ignoring higher-order neighbor hoppings—produces two dispersive bands (black curve in Figure 1C). When hopping is constrained within the 1D system, introducing an additional, relatively small hopping term $\boldsymbol{t_2}$ (Figure 1A), intriguingly flattens the dispersive band, as displayed by the red curve in Figure 1C. On the other hand, 1D Lieb lattice (Figure 1B) takes the effective Hamiltonian form:

$$\boldsymbol{H} = \begin{pmatrix} 0 & t_1 & 0 & 2t_2\cos(k/2) & 0 \\ t_1 & 0 & t_1 & 0 & 0 \\ 0 & t_1 & 0 & 0 & 2t_2\cos(k/2) \\ 2t_2\cos(k/2) & 0 & 0 & 0 & 0 \\ 0 & 0 & 2t_2\cos(k/2) & 0 & 0 \end{pmatrix}.$$

If $\boldsymbol{t_1 = t_2}$, it features an ideal 1D Lieb lattice, which holds a flat band at $\boldsymbol{E = 0}$ eV as shown in Figure 1D. Notably, the competition between $\boldsymbol{t_1}$ and $\boldsymbol{t_2}$ can modulate the dispersion of flat bands in this Lieb system.

In this work, the 1D PNRs with pentagonal nature, penta-hexa-PNRs and penta-dodeca-PNRs, can provide the platform to realize a set of robust arrangements of lD zigzag (with a relatively small hopping $\boldsymbol{t_2}$, Figure 1E) and Lieb lattices (Figure 1F), that can flexibly generate the flat bands (as discussed in Figure 1A to 1D). More importantly, the electronic band structures based on first-principles calculations of freestanding penta-hexa-PNRs and penta-dodeca-PNRs are plotted in Figure 1G and 1H, wherein the flat bands are clearly resolved. For the penta-hexa-PNRs, we further provide TB band structures with fitted hopping parameters in Figure 1G, which exhibits a distinct flat band that aligns well with the flat band obtained in the first-principles calculations. Notably, in the phosphorous materials systems, when comparing with the 1D zigzag and Lieb TB models, the position and the dispersion of the flat bands can be adjusted by atomic on-site potentials and higher-order neighboring hoppings, respectively. Moreover, the multi-orbital characteristics of phosphorous will give rise to more rich flat bands.

### 2.2. Synthesis and Electronic Structures of Penta-Hexa-PNRs

Black phosphorene, a mono-elemental 2D material with outstanding and highly directional properties, has emerged with increasing interest recently.[42-44] Intriguingly, the weak bonding of phosphorous element holds the promises of rich tunability, i.e., a variety of phosphorene allotropes including their corresponding nanoribbons have been theoretically predicted with extraordinary properties.[45-49] Thus, we chose phosphorous as the candidate element to grow the 1D system. To dictate the symmetry of the as-grown phosphorous nanostructures into a 1D material system, we opted for a specific interfacial interaction of Ag(110) surface, which exhibits anisotropy along the $x$ and $y$ directions for the preferential 1D growth [the discussion on 1D systems of phosphorous on surface with strong interfacial interaction,[50, 51] such as Au(110) and Cu(110)]. Here, we report the direct synthesis of 1D PNRs with pentagonal nature on Ag(110) via the molecular beam epitaxy (MBE).

Figure 2A illustrates the schematic diagram detailing the fabrication process of 1D PNRs, utilizing bulk black phosphorous as precursor under ultrahigh vacuum (UHV) conditions. In this controlled experiment, the evaporated phosphorous primarily consists of $P_4$ molecules.[52] Intriguingly, the precise control of substrate temperatures, specifically at low temperature (363 K) and high temperature (473 K), enables the well-controlled formation of two phases of PNRs with high uniformity. This is evidenced by the large-scale and close-up STM images of the two PNRs phases in Figure 2B-2E, revealing unidirectionally oriented 1D nanostripe arrays covering the surface. The formation of 1D PNRs involves not only the contribution of Ag(110) surface with anisotropy, but also relies significantly on the stabilizing effect introduced by the pentagonal nature of PNRs, as elaborated below. At a relatively higher substrate temperature, phosphorous atoms exhibit increased activity, facilitating surface diffusion. This higher surface diffusion length of phosphorous leads to the formation of 1D PNRs with a relatively larger unit cell, as shown in Figure 2C, 2E. These PNRs can cover almost the entire terrace of the Ag(110) substrate (Figure 2B, 2D), which not only ensures the feasibility of atomically resolved STM/STS measurements but also guarantees high-quality ARPES measurements.

Figure 3A displays a representative STM topographic image of PNRs prepared at low temperature (363 K). Upon detailed examination of the PNRs structure and the lattice constants of Ag(110) surface, it is revealed that the unit cell of the PNRs is well matched with the (2 × 5) supercell of Ag(110). Nonetheless, the precise determination of internal bond structures in the PNRs using STM is limited. Recent advancements in BR-STM enable the single-bond resolution imaging of the backbone of molecular and surface nanostructures using a carbon-monoxide (CO) functionalized STM tip. The deflection of the CO molecule arising from Pauli repulsion over the electron-rich areas effectively modulates the overall tunnelling conductance of the tip–sample junction, leading to the appearance of sharp line features over the positions of chemical bonds in the tunnelling current image.[53-55] Therefore, we performed BR-STM imaging to better resolve atomic structures of the PNRs at the single-bond level. The corresponding BR-STM image of this PNRs is displayed in Figure 3B, 3C. The observed bright triangles in STM are resolved into typical phosphorous pentamers, highlighted by red pentagon, comprising three relatively buckled up atoms (marked as a, b, c) and two relatively buckled down atoms (marked as d, e,). Besides, some electron clouds features are observed between these phosphorous pentamers, which can be ascribed to the single phosphorous atom underneath with buckled down configuration. Here, this bonding configuration with different-heights-buckling stems from the intrinsic $sp^3$ hybridization of phosphorous. Thus, the atomic structure of PNRs composed of pentagons and hexagons has been proposed in Figure 3D, namely penta-hexa-PNRs, and the simulated STM image (Figure 3E) agrees well with the experimental observations.

Experimental construction of such specific structure inspires us to further investigate its electronic structures. Here, the high uniformity of the penta-hexa-PNRs grown on Ag(110) enables the high-resolution ARPES measurements combined with low-energy electron diffraction (LEED) measurements. The LEED pattern of the penta-hexa-PNRs demonstrates a 2 × 5 superstructure with respect to Ag(110), in agreement with STM analysis. This also indicates the high uniformity of our samples. The ARPES spectra along $\Gamma-\overline{Y}-\Gamma$ are displayed in Figure 3F. Clearly, the ARPES spectra show a fine flat band at ~−2.0 eV, which is absent in pristine Ag(110). Thus, the fine flat band can be ascribed to the contribution of penta-hexa-PNRs. The ARPES spectra cuts along PNRs direction that are away from Γ point clearly demonstrate the robustness of flat bands. As for ARPES spectra cuts perpendicular to PNRs direction, only several trivial atomic flat bands emerge. Both strongly suggest the flat band only disperses in one direction, i.e., along the PNRs, and there is nearly no interaction between the PNRs, which is consistent with its 1D nature. It should also be noted that the linear dispersion bands with Dirac-cone-like features have been observed at ~−1.2 eV of the ARPES spectra, as indicated by the black arrow in Figure 3F, which is due to band folding induced by the PNRs periodicity,[56] as will be further illustrated in the following theoretical calculations.

To understand the observed band structure, we first calculated the electronic band structure of the penta-hexa-PNRs on one layer Ag(110) using the first-principles calculations in Figure 3G, see Experimental Methods. Obviously, one flat band exists below the Fermi level (~−1.5 eV, indicated by the red arrow), which is correlated with the flat band observed in ARPES measurements. In experiment, the penta-hexa-PNRs were prepared at a relatively low temperature (363 K), which results in the unwanted phosphorous on the surface (Figure 2B). This electron doping effect of the redundant phosphorous will adjust the Fermi level and leads to the flat band energy level difference between the experimental and calculation results. To further understand the linear dispersion bands with Dirac-cone-like features, we calculated the unfolded effective band structures of penta-hexa-PNRs/Ag(110) in the first BZ of Ag(110).[56] The unfolding band structures along $\Gamma-\overline{Y}-\Gamma$ of Ag(110) are shown in Figure 3H, wherein the Dirac-cone-like bands (located at ~−0.7 eV, in qualitative agreement with ARPES results considering the electron doping effect) are produced at the $\overline{Y}$ point of Ag(110), which is absent in the DFT calculations in Figure 3G. Importantly, the identified flat band is intrinsically from the penta-hexa-PNRs, as revealed by comparing the band structures of freestanding penta-hexa-PNRs and penta-hexa-PNRs/Ag(110).

To further investigate the electronic structures of the penta-hexa-PNRs, confirm the flat bands experimentally at the atomic scale and understand the origin of flat bands, we performed the STS measurements, in which the d$I$/d$V$ point spectrum probes the local electronic density of states (DOS), and d$I$/d$V$ map can reveal the spatial distribution of DOS at specific energy. As expected for the flat band with the macroscopically degenerate eigenstates, the measured spectroscopy can display the pronounced peaks with narrow width near the flat band energy region. The flat band is experimentally reflected by the sharp peak at −1.44 V in the d$I$/d$V$ spectrum in Figure 3I. The d$I$/d$V$ spectrum can be perfectly reproduced by the partial density of states (PDOS) of penta-hexa-PNRs on one layer Ag(110), wherein an obvious peak arising from the flat band can be observed (Figure 3J). Furthermore, the d$I$/d$V$ map under constant height mode reveals the spatial distribution associated with the electronic states in real space. Figure 3K is the d$I$/d$V$ map acquired at −1.44 V, demonstrating the spatial distribution of the flat band, which presents a typical zigzag lattice signature. Importantly, the corresponding theoretical d$I$/d$V$ map in Figure 3L using the probe particle STM (PP-STM) simulation (see Experimental Methods) exhibits good agreement with experimental results as well.

Notably, it is indeed challenging to ensure that all the rigid conditions for achieving perfect flat bands are fully met in real materials, i.e., the position and the dispersion of the flat bands can be adjusted by atomic on-site potentials and higher-order neighboring hoppings. Here, by confining the hoppings to 1D system, the constructed 1D flat bands exhibit minimal dispersion. The corresponding STS measurements demonstrate that the dominant hoppings follow a zigzag pattern for penta-hexa-PNRs. Therefore, we have demonstrated the successful construction of flat band in penta-hexa-PNRs, derived from electronic 1D zigzag lattice.

**2.3. Synthesis and Electronic Structures of Penta- Dodeca-PNRs**

To further explore how general the advantages of constructing flat bands in 1D system could be, we also constructed 1D flat bands in penta-dodeca-PNRs prepared at higher annealing temperature (Figure 4A), whose structure was carefully analyzed in terms of the lattice constants of Ag(110) surface, which reveals that the unit cell of the PNRs is well matched with the (2 × 8) supercell of Ag(110). Figure 4A is close-up STM image of the new PNRs phase. Similarly, we also performed the BR-STM measurements to reveal the atomic structure. The corresponding BR-STM images in Figure 4 (B and C), reveal that the interconnected head-to-head phosphorous pentamers (marked as the red pentagonal) are linked by single buckled-down phosphorous atoms, wherein the pentagonal phosphorous also contains three buckled-up (marked as a, b, c) and two buckled-down phosphorous atoms (marked as d, e). Notably, a bright feature can also be observed within the backbone of PNRs, where no obvious chemical bonds are formed with the neighboring atoms. This can be assigned as a single silver adatom since the phosphorous atom will form strong bond with the neighboring phosphorous atoms. Thus, the atomic structure of PNRs containing pentagons and dodecagons, namely penta-dodeca-PNRs, has been proposed in Figure 4D, which is supported by the excellent agreement between the simulated STM image (Figure 4E) and experimentally observed STM image. Therefore, the structure of penta-dodeca-PNRs has been solidly confirmed.

Intriguingly, the penta-dodeca-PNRs feature a typical 1D Lieb lattice shape, a square depleted lattice, which predicts the formation of flat band (Figure 1B, 1D). Such structure has inspired us to further investigate its electronic structure. The LEED measurements were also performed to determine the high uniformity of our sample before the ARPES measurements, wherein a clear 2 × 8 superstructure with respect to Ag(110) is observed, consistent with STM measurements. Here, the ARPES spectra were cut along the blue line of the inset in Figure 4F that are away from the $\bar{\Gamma}$ point, since the bulk bands of Ag(110) have strong spectral weights near the $\bar{\Gamma}$ point and overlap with the bands of the PNRs. Indeed, as revealed by the ARPES measurements of the penta-dodeca-PNRs (Figure 4F), one obvious flat band with tiny dispersion has been observed at ~−2.0 eV, indicated by the red arrow. The flat band can be successfully produced by the DFT calculations of the penta-dodeca-PNRs on one layer Ag(110), which emerges at the ~−1.8 eV in the band structures (red arrow in Figure 4G), in good agreement with the ARPES results. Similarly, it is important that the identified flat band is intrinsic to the penta-dodeca-PNRs. Here, the flat band becomes more dispersive in the presence of the Ag substrate compared to the freestanding penta-dodeca-PNRs. This occurs because (1) the buckled-down phosphorus atoms at pentagon junctions couple with the substrate, and (2) the adsorbed silver atoms hybridize with bridging phosphorus atoms along the chain direction. In experiment, the penta-dodeca-PNRs were prepared at a relatively high temperature comparing with that of penta-hexa-PNRs. Under this case, the redundant phosphorous atoms cannot reside on the surface, and consequently no electron doping effect comes into play.

The observed flat bands also feature representative 1D characteristics (ARPES measurements along the penta-dodeca-PNRs direction and ARPES measurements cutting perpendicular to PNRs direction). Furthermore, the linear dispersion bands with Dirac-cone-like features emerge at ~−1.2 eV, indicated by the black arrow, which can be assigned as the band folding effect induced by the PNRs periodicity as well.[56] This is also well supported by the calculated unfolding band structures (Figure 4H). It is worth noting that the tight-binding band structure of the 1D Lieb lattice based on single orbital model with hopping strengths of $t_1 = t_2 = 1$ (Figure 1D) exhibits only one flat band. In this 1D penta-dodeca-PNRs system, on the one hand, the competition between $t_1$ and $t_2$ can modulate the dispersion of flat bands in this Lieb system (Figure 4F); on the other hand, the valence electrons of P atoms have multi orbitals and their corresponding hopping strengths and the on-site energy may vary, thus yielding two sets of flat bands (the other one locates at the +2.5 eV). Detailed analysis of such flat bands will be illustrated in the next section.

### 2.4. Multiple Flat Bands in the Penta- Dodeca-PNRs

The STS measurements enable the investigation of flat bands at the atomic scale, but are not limited to the flat bands only below Fermi level comparing with ARPES measurements. Figure 5 (A and B) displays the $\mathrm{d}I/\mathrm{d}V$ point spectrum of the penta-dodeca-PNRs, wherein three pronounced peaks can be clearly observed, located at −1.70 V, +1.95 V and +2.46 V, respectively. We also calculated the PDOS for the penta-dodeca-PNRs on one layer Ag(110) (Figure 5C), where a series of peaks arising from the electronic band structure are consistent with the experimental d$I$/d$V$ spectra, i.e., the energy peaks of −1.70 V and +2.46 V are phosphorous DOS dominant while that of +1.95 V is mostly silver related.

By analyzing the d$I$/d$V$ maps and the corresponding STM images, the d$I$/d$V$ maps collected at energies of −1.70 V and +2.46 V in Figure 5D show that their spatial distribution is mainly derived from the penta-dodeca-PNRs and form a 1D Lieb lattice, wherein the electron hopping strength along the horizontal direction is relatively larger than that of along the vertical direction at −1.70 V while conversely at +2.46 V. On the other hand, the corresponding map of the peak located at +1.95 V is mainly distributed in the central silver atoms and single phosphorous bridging atoms, which indicates the peak is related to silver contribution. Thus, the energy peaks of −1.70 V and +2.46 V can be attributed to the two emerging flat bands ($FB_1$ and $FB_2$) of the penta-dodeca-PNRs on Ag(110) in Figure 4G, with both electron hoppings following a 1D Lieb lattice. The competing hoppings result in the tiny dispersion of the observed flat band in Figure 4F. Furthermore, the theoretical d$I$/d$V$ maps of the penta-dodeca-PNRs on Ag(110) at energies peaks corresponding to the $FB_1$ and $FB_2$ match well with the experimental patterns, as well for the peak of +1.95 V, as displayed in Figure 5E. Therefore, the agreement between the DFT calculations and the experimental observations confirms the two flat bands in the penta-dodeca-PNRs, which indeed originates from the 1D Lieb lattice with relatively alternative hopping strengths.

Fabricating real solid materials with ideal flat bands is crucial for the fundamental many-body physics research and applications. This remains as a significant challenge because the electron hoppings along specific geometries are always required to induce destructive interference. In our material systems, by reducing the dimensionality, the electron hoppings have been successfully confined into 1D systems with high tunability, wherein the supporting substrate plays a significant impact on the 1D PNRs epitaxial growth. The specific symmetries in these 1D PNRs with pentagonal nature can be analogous to 1D zigzag and Lieb lattices, respectively, thus producing ideal 1D flat bands. It is worth noting that the magic pentagonal phosphorus also plays a critical role in stabilizing the two phases of 1D PNRs, as the $sp^3$ hybridization of phosphorous usually tends to form a hexagonal shape and extends into 2D system. More importantly, the identified flat bands are intrinsically from the PNRs, as revealed by comparing the band structures of freestanding PNRs and PNRs/Ag(110). The electron hopping within these PNRs still follows the line graphs of 1D zigzag/Lieb lattices even with the substrate taken into consideration, as also revealed by the good agreements between DFT calculations and d$I$/d$V$ mapping.

**3. Conclusion**

In summary, we have demonstrated the direct synthesis of two phases of atomically precise 1D PNRs with pentagonal nature on Ag(110) surface. Our results provide a simple but general strategy for constructing flat bands in 1D systems and reveal that the emergent 1D flat bands in penta-hexa-PNRs and penta-dodeca-PNRs are originated from the electronic 1D zigzag and Lieb lattices respectively. This approach can be utilized as a design criterion for constructing flat bands towards the desired topologies, with prosperous potential in achieving more general strongly correlated topological phenomena.

**4. Experimental Methods**

*Sample Preparation*: The clean Ag(110) single crystal was prepared by repeated cycles of $Ar^+$ sputtering ($5 \times 10^{-5}$ mbar, 20 min) with subsequent annealing (673 K, 20 min). Here, the black phosphorous was used as the precursor and deposited from Knudsen cell at 573 K onto heated Ag(110) surface under ultrahigh vacuum (UHV) conditions. The two phases of PNRs can be well prepared on Ag(110) by controlling the annealing substrate temperature. The penta-hexa-PNRs (penta-dodeca-PNRs) was prepared by depositing P atoms onto heated Ag(110) surface held at 363 K (473 K) for 15 min.

*STM/STS Measurements*: Experiments were performed in a Scienta Omicron LT-STM system under UHV conditions with a base pressure $< 2 \times 10^{-10}$ mbar. After the fabrication of PNRs, the sample was immediately transferred into the STM head for measurement. The tip was calibrated on the Au(111) substrate before measurements. All STM/STS measurements were conducted at 4.5 K, with the bias voltage applied to the sample. The BR-STM imaging was conducted in constant height mode in the Pauli repulsion regime with CO-functionalized tip. The CO-functionalized tip was prepared by picking up CO from Au(111) surface. The tip-sample distance with respect to an STM set point is indicated in the figure caption for each BR-STM image. STS measurements were conducted by using an internal lock-in amplifier with an amplitude of 20 mV at frequency of 693 Hz. All the d$I$/d$V$ maps were collected in the constant-height mode.

*ARPES Measurements*: ARPES experiments were performed in UHV conditions (base pressure, $< 2.5 \times 10^{-10}$ mbar) at 40 K with a SPECS PHOIBUS 150 electron energy analyzer and a helium discharge lamp (He Iα light, $h\nu = 21.2eV$). The quality of synthesized PNRs was confirmed by LEED before the ARPES measurements.

*First-Principles Calculations*: Our first-principles calculations were based on DFT as implemented in the Vienna *Ab initio* Simulation Package (VASP),[57, 58] using the projector augmented-wave method.[59] The generalized gradient approximation with the Perdew-Burke-Ernzerhof[60] realization was adopted for the exchange-correlation functional. The plane-wave cutoff energy was set to 500 eV with k-point mesh samplings of $15 \times 9 \times 1$ and $15 \times 5 \times 1$ for penta-hexa-PNRs and penta-dodeca-PNRs, respectively. The crystal structures with the fixed lattice constants ($a = 7.7$ Å, $b = 13.6$ Å for penta-hexa-PNRs; $a = 7.7$ Å, $b = 21.5$ Å for penta-dodeca-PNRs;), placed on Ag(110) slab of 6 layers, allowing the atoms on Ag to optimize until the forces on the atoms were less than 0.001 eV/Å, consistent with previous studies.[61, 62] To avoid unnecessary interactions between the monolayer and its periodic images, the vacuum layer was set to 15 Å. Calculations on freestanding penta-hexa-PNRs/penta-dodeca-PNRs, as well as those on PNSs with a single layer of Ag(110) substrate, were performed based on the previously optimized structures.

*dI/dV Maps Simulations*: All the constant-height d$I$/d$V$ maps simulations were calculated using the PP-STM code.[63] The CO tip was positioned around 4.5 Å above the sample. To achieve the best agreement with the experimental d$I$/d$V$ maps, different ratios of $s$ and $p_{xy}$ orbitals on the oxygen was adopted (10% $s$ and 90% $p_{xy}$ for the Figure 3L in the penta-hexa-PNRs and the left panel of Figure 5E in the penta-dodeca-PNRs; 15% $s$ and 85% $p_{xy}$ for the middle and right panel of Figure 5E in the penta-dodeca-PNRs), which agree well with previously published results.[63-65] The eigenenergies and the linear combination of atomic orbitals (LCAO) coefficients of PNRs, which are needed for the aforementioned d$I$/d$V$ maps simulations, were calculated using the FHI-aims code with the default tight basis set.[66] Noting that the PP-STM code can only read the Γ−point data, the $5 \times 3$ and $5 \times 2$ supercells were thus adopted for the penta-hexa-PNRs and penta-dodeca-PNRs, respectively, and the BZ was represented by the Γ point only.

**Acknowledgements**

This work was financially supported by Natural Science Foundation of China (62274118, 12304199, 12004278, 11974391, 62375052, 12304481 U2032204), Singapore National Research Foundation Investigatorship under Grant No. NRF-NRFI08-2022-0009, SUSTech-NUS Joint Research Program, and Singapore MOE AcRF Tier 2 Grants of MOE-T2EP50220-0001 and MOE-T2EP10221-0005, Tier 3 Award MOE2018-T3-1-002. B.F. acknowledges support from the Ministry of Science and Technology of China (Grant No. 2018YFE0202700), the International Partnership Program of Chinese Academy of Sciences (Grant No. 112111KYSB20200012), and the Strategic Priority Research Program of Chinese Academy of Sciences (Grants No. XDB33030100). S.S. acknowledges support from Science and Technology Commission of Shanghai Municipality, the Shanghai Venus Sailing Program (Grant No. 23YF1412600). J.L.Z. acknowledges support from National Key Research and Development Program of China (2022YFB4400100), Natural Science Foundation of Jiangsu Province under BK20210199 and Jiangsu Provincial Double-Innovation Doctor Program JSSCBS20210141. J.S. acknowledges the support from Agency for Science, Technology and Research (A*STAR) Advanced Manufacturing & Engineering (AME) Young Individual Research Grant (YIRG) A2084c0171.

**Author Contributions**

S.S. and W.C. conceived the experiments. S.S., J.S. and X.P. performed the MBE growth and STM/STS measurements under the guidance of J.L.Z., J.L. and W.C.; J.-Y.Y. and T. Y. performed the *Ab initio* calculations with the assistance of Y.P.F.; Z.C. and D.G. performed ARPES measurements with the assistance of L.C., K.W. and B.F.; Y.W., J.G., Y.H. and S.D. helped acquire the experimental data. S.S., J.-Y.Y., J.L.Z., B.F., A.T.S.W. and W.C. analyzed the experimental and theoretical data. S.S. J.-Y.Y., and W.C. wrote the manuscript. All authors discussed the results and commented on the manuscript. These authors contributed equally: S. S., J. Y., Z. C. and J. S.

References

[1] R. Bistritzer, A. H. MacDonald, *Proc Natl Acad Sci U S A* **2011**, 108, 12233.
[2] M. Imada, M. Kohno, *Physical Review Letters* **2000**, 84, 143.
[3] D. N. Sheng, Z.-C. Gu, K. Sun, L. Sheng, *Nature Communications* **2011**, 2, 389.
[4] K. Sun, Z. Gu, H. Katsura, S. Das Sarma, *Phys Rev Lett* **2011**, 106, 236803.
[5] J. W. Rhim, K. Kim, B. J. Yang, *Nature* **2020**, 584, 59.
[6] Y. Hwang, J.-W. Rhim, B.-J. Yang, *Nature Communications* **2021**, 12, 6433.
[7] D. Leykam, A. Andreanov, S. Flach, *Advances in Physics: X* **2018**, 3, 1473052.
[8] L. Balents, C. R. Dean, D. K. Efetov, A. F. Young, *Nature Physics* **2020**, 16, 725.
[9] J.-W. Rhim, B.-J. Yang, *Advances in Physics: X* **2021**, 6.
[10] Y. Cao, V. Fatemi, S. Fang, K. Watanabe, T. Taniguchi, E. Kaxiras, P. Jarillo-Herrero, *Nature* **2018**, 556, 43.
[11] Y. Cao, V. Fatemi, A. Demir, S. Fang, S. L. Tomarken, J. Y. Luo, J. D. Sanchez-Yamagishi, K. Watanabe, T. Taniguchi, E. Kaxiras, R. C. Ashoori, P. Jarillo-Herrero, *Nature* **2018**, 556, 80.
[12] S. Lisi, X. Lu, T. Benschop, T. A. de Jong, P. Stepanov, J. R. Duran, F. Margot, I. Cucchi, E. Cappelli, A. Hunter, A. Tamai, V. Kandyba, A. Giampietri, A. Barinov, J. Jobst, V. Stalman, M. Leeuwenhoek, K. Watanabe, T. Taniguchi, L. Rademaker, S. J. van der Molen, M. P. Allan, D. K. Efetov, F. Baumberger, *Nature Physics* **2021**, 17, 189.
[13] Z. Li, J. Zhuang, L. Wang, H. Feng, Q. Gao, X. Xu, W. Hao, X. Wang, C. Zhang, K. Wu, S. X. Dou, L. Chen, Z. Hu, Y. Du, *Science Advances* **2018**, 4, eaau4511.
[14] Z. Zhang, Y. Wang, K. Watanabe, T. Taniguchi, K. Ueno, E. Tutuc, B. J. LeRoy, *Nature Physics* **2020**, 16, 1093.
[15] H. Li, S. Li, M. H. Naik, J. Xie, X. Li, J. Wang, E. Regan, D. Wang, W. Zhao, S. Zhao, S. Kahn, K. Yumigeta, M. Blei, T. Taniguchi, K. Watanabe, S. Tongay, A. Zettl, S. G. Louie, F. Wang, M. F. Crommie, *Nature Materials* **2021**, 20, 945.

[16] G.-B. Jo, J. Guzman, C. K. Thomas, P. Hosur, A. Vishwanath, D. M. Stamper-Kurn, *Physical Review Letters* **2012**, 108, 045305.
[17] S. Mukherjee, A. Spracklen, D. Choudhury, N. Goldman, P. Öhberg, E. Andersson, R. R. Thomson, *Physical Review Letters* **2015**, 114, 245504.
[18] F. Baboux, L. Ge, T. Jacqmin, M. Biondi, E. Galopin, A. Lemaître, L. Le Gratiet, I. Sagnes, S. Schmidt, H. E. Türeci, A. Amo, J. Bloch, *Physical Review Letters* **2016**, 116, 066402.
[19] M. R. Slot, T. S. Gardenier, P. H. Jacobse, G. C. P. van Miert, S. N. Kempkes, S. J. M. Zevenhuizen, C. M. Smith, D. Vanmaekelbergh, I. Swart, *Nat Phys* **2017**, 13, 672.
[20] M. N. Huda, S. Kezilebieke, P. Liljeroth, *Physical Review Research* **2020**, 2, 043426.
[21] N. Regnault, Y. Xu, M.-R. Li, D.-S. Ma, M. Jovanovic, A. Yazdani, S. S. P. Parkin, C. Felser, L. M. Schoop, N. P. Ong, R. J. Cava, L. Elcoro, Z.-D. Song, B. A. Bernevig, *Nature* **2022**, 603, 824.
[22] D. Călugăru, A. Chew, L. Elcoro, Y. Xu, N. Regnault, Z.-D. Song, B. A. Bernevig, *Nature Physics* **2021**, 18, 185.
[23] H. Liu, S. Meng, F. Liu, *Physical Review Materials* **2021**, 5.
[24] H. Liu, G. Sethi, S. Meng, F. Liu, *Physical Review B* **2022**, 105.
[25] W. Maimaiti, A. Andreanov, H. C. Park, O. Gendelman, S. Flach, *Physical Review B* **2017**, 95, 115135.
[26] W. Maimaiti, S. Flach, A. Andreanov, *Physical Review B* **2019**, 99, 125129.
[27] W. Maimaiti, A. Andreanov, S. Flach, *Physical Review B* **2021**, 103, 165116.
[28] J.-W. Rhim, B.-J. Yang, *Physical Review B* **2019**, 99, 045107.
[29] J. Ma, J.-W. Rhim, L. Tang, S. Xia, H. Wang, X. Zheng, S. Xia, D. Song, Y. Hu, Y. Li, B.-J. Yang, D. Leykam, Z. Chen, *Physical Review Letters* **2020**, 124, 183901.
[30] W. Jiang, S. Zhang, Z. Wang, F. Liu, T. Low, *Nano Letters* **2020**, 20, 1959.
[31] S. Sun, S. Zhao, Y. Z. Luo, X. Gu, X. Lian, A. Tadich, D.-C. Qi, Z. Ma, Y. Zheng, C. Gu, J. L. Zhang, Z. Li, W. Chen, *Nano Letters* **2020**, 20, 5583.
[32] L. Ye, M. Kang, J. Liu, F. von Cube, C. R. Wicker, T. Suzuki, C. Jozwiak, A. Bostwick, E. Rotenberg, D. C. Bell, L. Fu, R. Comin, J. G. Checkelsky, *Nature* **2018**, 555, 638.
[33] Z. Lin, J.-H. Choi, Q. Zhang, W. Qin, S. Yi, P. Wang, L. Li, Y. Wang, H. Zhang, Z. Sun, L. Wei, S. Zhang, T. Guo, Q. Lu, J.-H. Cho, C. Zeng, Z. Zhang, *Physical Review Letters* **2018**, 121, 096401.
[34] J.-X. Yin, S. S. Zhang, G. Chang, Q. Wang, S. S. Tsirkin, Z. Guguchia, B. Lian, H. Zhou, K. Jiang, I. Belopolski, N. Shumiya, D. Multer, M. Litskevich, T. A. Cochran, H. Lin, Z. Wang, T. Neupert, S. Jia, H. Lei, M. Z. Hasan, *Nature Physics* **2019**, 15, 443.
[35] M. Kang, L. Ye, S. Fang, J.-S. You, A. Levitan, M. Han, J. I. Facio, C. Jozwiak, A. Bostwick, E. Rotenberg, M. K. Chan, R. D. McDonald, D. Graf, K. Kaznatcheev, E. Vescovo, D. C. Bell, E. Kaxiras, J. van den Brink, M. Richter, M. Prasad Ghimire, J. G. Checkelsky, R. Comin, *Nature Materials* **2020**, 19, 163.
[36] M. Kang, S. Fang, L. Ye, H. C. Po, J. Denlinger, C. Jozwiak, A. Bostwick, E. Rotenberg, E. Kaxiras, J. G. Checkelsky, R. Comin, *Nature Communications* **2020**, 11, 4004.

[37] Z. Liu, M. Li, Q. Wang, G. Wang, C. Wen, K. Jiang, X. Lu, S. Yan, Y. Huang, D. Shen, J.-X. Yin, Z. Wang, Z. Yin, H. Lei, S. Wang, *Nature Communications* **2020**, 11, 4002.
[38] M. Li, Q. Wang, G. Wang, Z. Yuan, W. Song, R. Lou, Z. Liu, Y. Huang, Z. Liu, H. Lei, Z. Yin, S. Wang, *Nature Communications* **2021**, 12, 3129.
[39] H. Chen, H. Yang, B. Hu, Z. Zhao, J. Yuan, Y. Xing, G. Qian, Z. Huang, G. Li, Y. Ye, S. Ma, S. Ni, H. Zhang, Q. Yin, C. Gong, Z. Tu, H. Lei, H. Tan, S. Zhou, C. Shen, X. Dong, B. Yan, Z. Wang, H. J. Gao, *Nature* **2021**, 599, 222.
[40] Y. X. Jiang, J. X. Yin, M. M. Denner, N. Shumiya, B. R. Ortiz, G. Xu, Z. Guguchia, J. He, M. S. Hossain, X. Liu, J. Ruff, L. Kautzsch, S. S. Zhang, G. Chang, I. Belopolski, Q. Zhang, T. A. Cochran, D. Multer, M. Litskevich, Z. J. Cheng, X. P. Yang, Z. Wang, R. Thomale, T. Neupert, S. D. Wilson, M. Z. Hasan, *Nat Mater* **2021**, 20, 1353.
[41] L. Nie, K. Sun, W. Ma, D. Song, L. Zheng, Z. Liang, P. Wu, F. Yu, J. Li, M. Shan, D. Zhao, S. Li, B. Kang, Z. Wu, Y. Zhou, K. Liu, Z. Xiang, J. Ying, Z. Wang, T. Wu, X. Chen, *Nature* **2022**, 604, 59.
[42] L. Li, Y. Yu, G. J. Ye, Q. Ge, X. Ou, H. Wu, D. Feng, X. H. Chen, Y. Zhang, *Nature Nanotechnology* **2014**, 9, 372.
[43] J. Kim, S. S. Baik, S. H. Ryu, Y. Sohn, S. Park, B.-G. Park, J. Denlinger, Y. Yi, H. J. Choi, K. S. Kim, *Science* **2015**, 349, 723.
[44] H. Liu, A. T. Neal, Z. Zhu, Z. Luo, X. Xu, D. Tománek, P. D. Ye, *ACS Nano* **2014**, 8, 4033.
[45] A. Carvalho, A. S. Rodin, A. H. Castro Neto, *Europhysics Letters* **2014**, 108, 47005.
[46] Q. Wu, L. Shen, M. Yang, Y. Cai, Z. Huang, Y. P. Feng, *Physical Review B* **2015**, 92, 035436.
[47] Z. Zhu, D. Tománek, *Physical Review Letters* **2014**, 112, 176802.
[48] J. Guan, Z. Zhu, D. Tománek, *Physical Review Letters* **2014**, 113, 046804.
[49] M. Wu, H. Fu, L. Zhou, K. Yao, X. C. Zeng, *Nano Letters* **2015**, 15, 3557.
[50] S. Sun, T. Yang, Z. Ma, H. Ding, S. Zhao, J. Hu, Q. Xu, X. Lian, C. Gu, Z. Li, *ACS Materials Letters* **2020**, 2, 873.
[51] J. L. Zhang, S. Zhao, S. Sun, T. Niu, X. Zhou, C. D. Gu, C. Han, K. D. Yuan, R. Guo, L. Wang, Z. Li, W. Chen, *Advanced Materials Interfaces* **2017**, 4, 1601167.
[52] R. W. Keyes, *Physical Review* **1953**, 92, 580.
[53] P. Hapala, G. Kichin, C. Wagner, F. S. Tautz, R. Temirov, P. Jelínek, *Physical Review B* **2014**, 90, 085421.
[54] G. D. Nguyen, H.-Z. Tsai, A. A. Omrani, T. Marangoni, M. Wu, D. J. Rizzo, G. F. Rodgers, R. R. Cloke, R. A. Durr, Y. Sakai, F. Liou, A. S. Aikawa, J. R. Chelikowsky, S. G. Louie, F. R. Fischer, M. F. Crommie, *Nature Nanotechnology* **2017**, 12, 1077.
[55] S. Song, N. Guo, X. Li, G. Li, Y. Haketa, M. Telychko, J. Su, P. Lyu, Z. Qiu, H. Fang, X. Peng, J. Li, X. Wu, Y. Li, C. Su, M. J. Koh, J. Wu, H. Maeda, C. Zhang, J. Lu, *Journal of the American Chemical Society* **2020**, 142, 13550.
[56] P. Gori, O. Pulci, F. Ronci, S. Colonna, F. Bechstedt, *Journal of Applied Physics* **2013**, 114.
[57] G. Kresse, J. Hafner, *Physical Review B* **1994**, 49, 14251.
[58] G. Kresse, J. Furthmüller, *Physical Review B* **1996**, 54, 11169.

[59] P. E. Blöchl, *Physical Review B* **1994**, 50, 17953.
[60] J. P. Perdew, K. Burke, M. Ernzerhof, *Physical Review Letters* **1996**, 77, 3865.
[61] Q. Zhong, L. Kong, J. Gou, W. Li, S. Sheng, S. Yang, P. Cheng, H. Li, K. Wu, L. Chen, *Physical Review Materials* **2017**, 1, 021001.
[62] Y. Wang, L. Kong, C. Chen, P. Cheng, B. Feng, K. Wu, L. Chen, *Advanced Materials* **2020**, 32, 2005128.
[63] O. Krejčí, P. Hapala, M. Ondráček, P. Jelínek, *Physical Review B* **2017**, 95, 045407.
[64] B. de la Torre, M. Švec, G. Foti, O. Krejčí, P. Hapala, A. Garcia-Lekue, T. Frederiksen, R. Zbořil, A. Arnau, H. Vázquez, P. Jelínek, *Physical Review Letters* **2017**, 119, 166001.
[65] Q. Fan, L. Yan, M. W. Tripp, O. Krejčí, S. Dimosthenous, S. R. Kachel, M. Chen, A. S. Foster, U. Koert, P. Liljeroth, J. M. Gottfried, *Science* **2021**, 372, 852.
[66] V. Blum, R. Gehrke, F. Hanke, P. Havu, V. Havu, X. Ren, K. Reuter, M. Scheffler, *Computer Physics Communications* **2009**, 180, 2175.

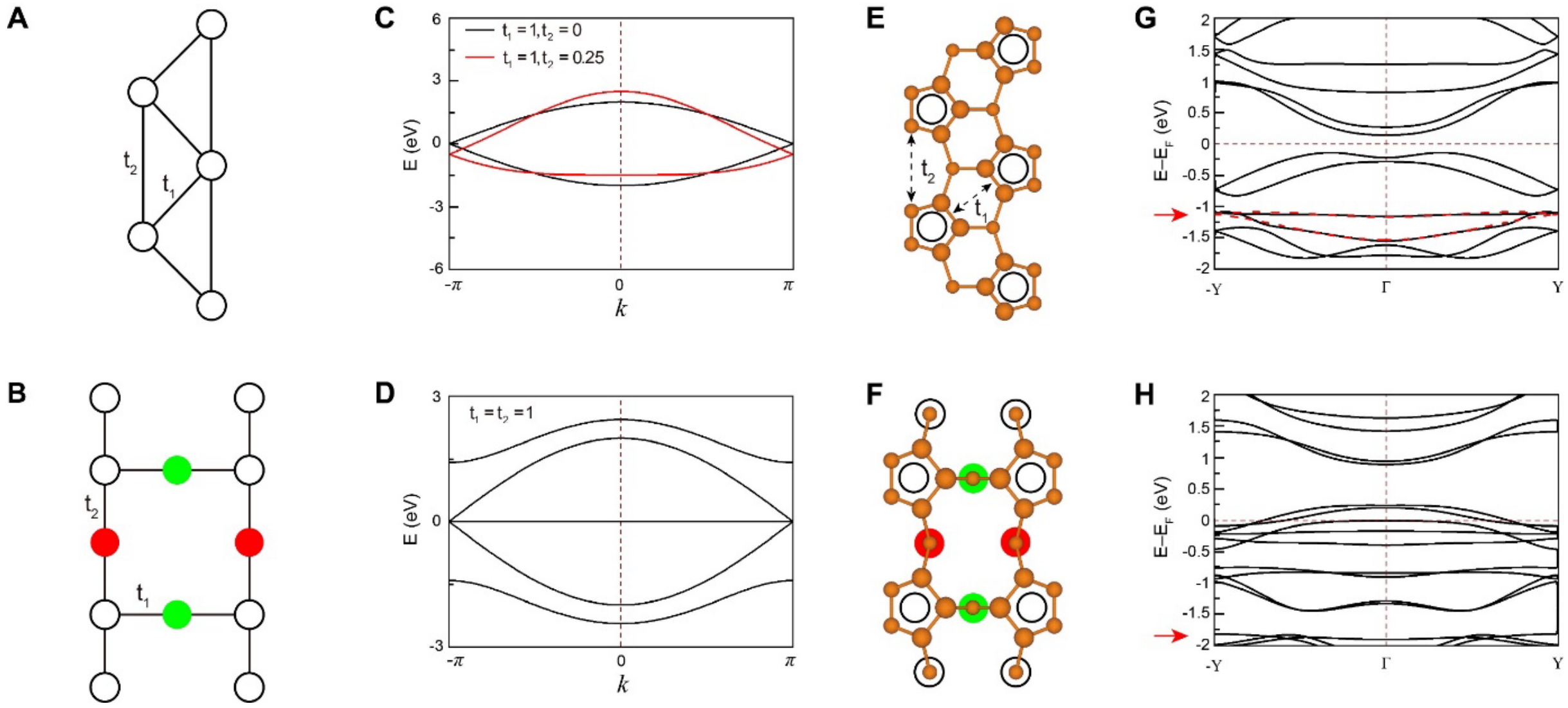

**Figure 1.** Flat bands construction in 1D systems. A, B) Schematic diagrams of 1D zigzag and Lieb lattices, where the hopping parameters are indicated by $t_1$ and $t_2$. C, D) Corresponding TB band structures of 1D zigzag lattice ($t_1 = 1,\ t_2 = 0$ black curve; $t_1 = 1,\ t_2 = 0.25$ red curve) and 1D Lieb lattice ($t_1 = t_2 = 1$). E, F) Schematics for the experimentally synthesized two phases of 1D PNRs with pentagonal nature: penta-hexa-PNRs (E) and penta-dodeca-PNRs (F), in which more buckled-up P atoms are represented by larger orange spheres. These two structures are equivalent to 1D zigzag lattice (E) and 1D Lieb lattice (F), respectively. G, H) Electronic band structures of freestanding penta-hexa-PNRs (G) and penta-dodeca-PNRs (H). The red dashed curve in (G) represents the TB band structures of 1D zigzag lattice with the fitting parameters: $t_1 = 0.091\ eV,\ t_2 = -0.057\ eV$, on-site energy$= -1.242\ eV$.

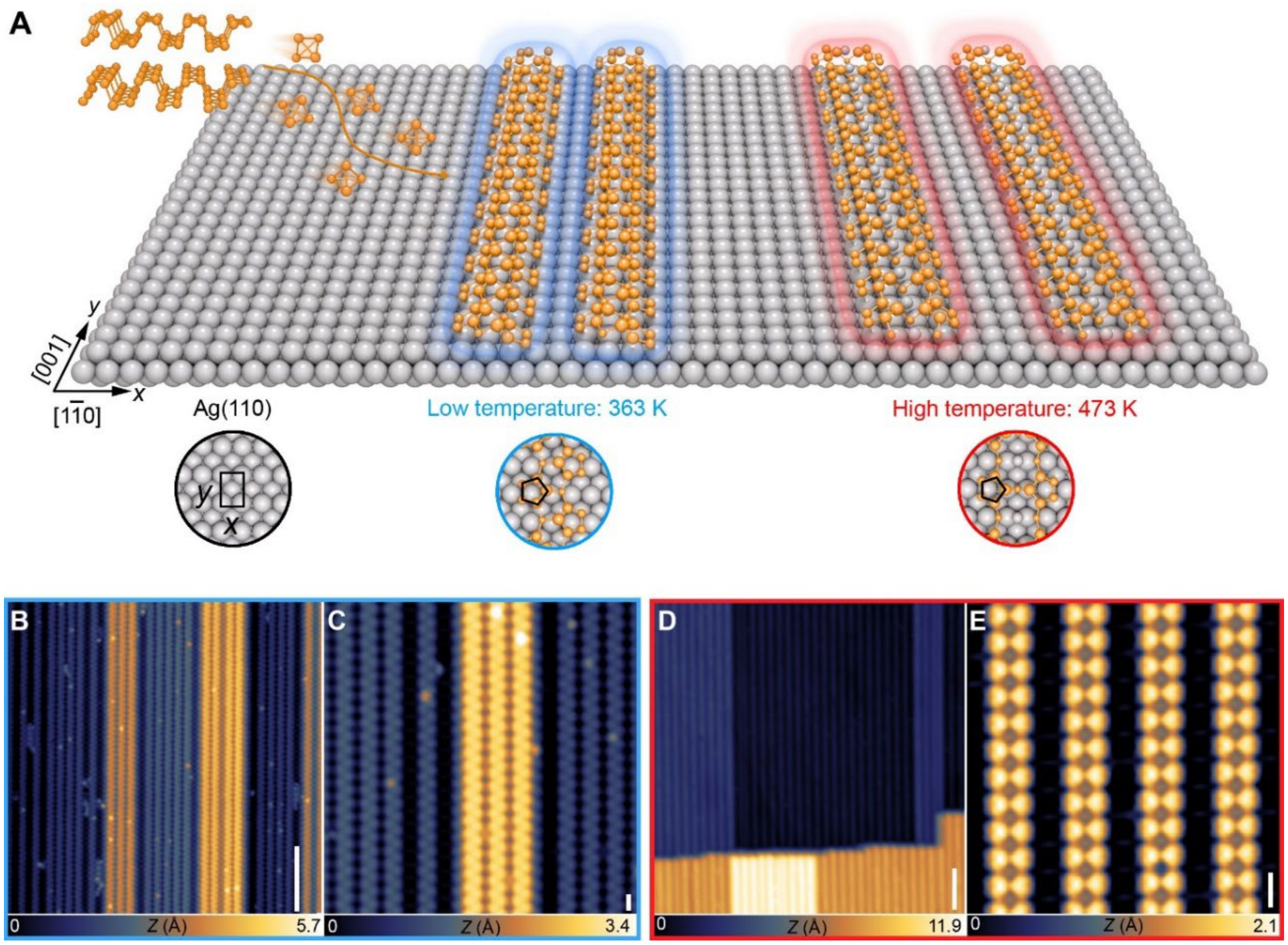

**Figure 2.** Bottom-up synthesis of 1D PNRs on Ag(110). A) Schematic diagram of the fabrication process on Ag(110) surface using bulk black phosphorous as the precursor, wherein the orange (gray) spheres correspond to P (Ag) atoms. B, C) Large-scale and close-up STM images of 1D PNRs on Ag(110), prepared at low temperature 363 K ($V$ = 1 V, $I$ = 0.3 nA). The relatively low substrate temperature resulted in the unwanted phosphorous on the surface, which mainly situated at the edges or the boundaries between the PNRs. D, E) Large-scale and close-up STM images of 1D PNRs on Ag(110), prepared at high temperature 473 K ($V$ = 1 V, $I$ = 0.1 nA; $V$ = 1 V, $I$ = 0.3 nA). Scale bar: 10 nm in (B, D); 1 nm in (C, E).

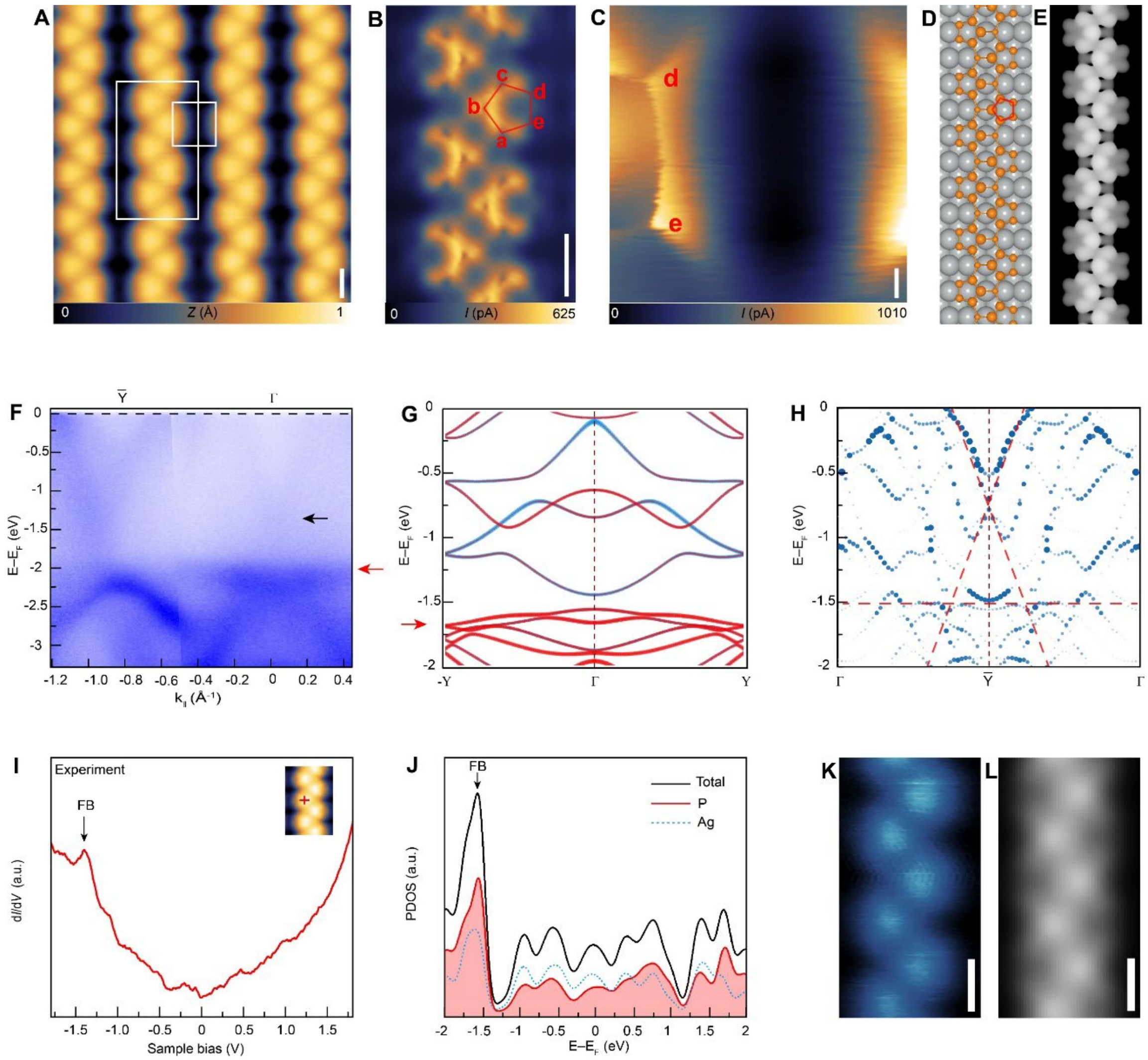

**Figure 3.** Atomic and electronic structures of penta-hexa-PNRs on Ag(110). A) Close-up STM image of the penta-hexa-PNRs on Ag(110) ($V$ = 100 mV, $I$ = 0.5 nA). B, C) BR-STM images of the penta-hexa-PNRs, taken at the white rectangles in panel (A) ($V$ = 1 mV, $\Delta Z$ = −0.6 Å; set point prior to turn off feedback, $V$ = 100 mV, $I$ = 50 pA). D, E) Top view of the penta-hexa-PNRs on Ag(110) and the corresponding simulated STM image, wherein the orange (gray) spheres correspond to P (Ag) atoms, in which more buckled-up P atoms are represented by larger orange spheres. F) ARPES spectra of penta-hexa-PNRs on Ag(110) along $\Gamma{-}\overline{Y}{-}\Gamma$. G) Electronic band structures of the penta-hexa-PNRs on one layer Ag(110) based on first-principles calculations, showing one flat band indicated by the red arrow. The red and blue dotted lines correspond to the contributions of P and Ag atoms, respectively. H) Calculated band structures projected to PNRs and topmost Ag layer along

$\Gamma-\overline{Y}-\Gamma$. The red dashed lines indicate the flat band and linear dispersion bands at the $\overline{Y}$ point. I) d$I$/d$V$ point spectrum with a signature of flat band peak, taken at the red cross in the inset. J) Calculated PDOS for the penta-hexa-PNRs on one layer Ag(110). K) d$I$/d$V$ map recorded at the peak of −1.44 V in (I). L) Calculated d$I$/d$V$ map at the energy of flat band in (J). Scale bar: 5 Å in (A, B, K, L); 0.5 Å in (C).

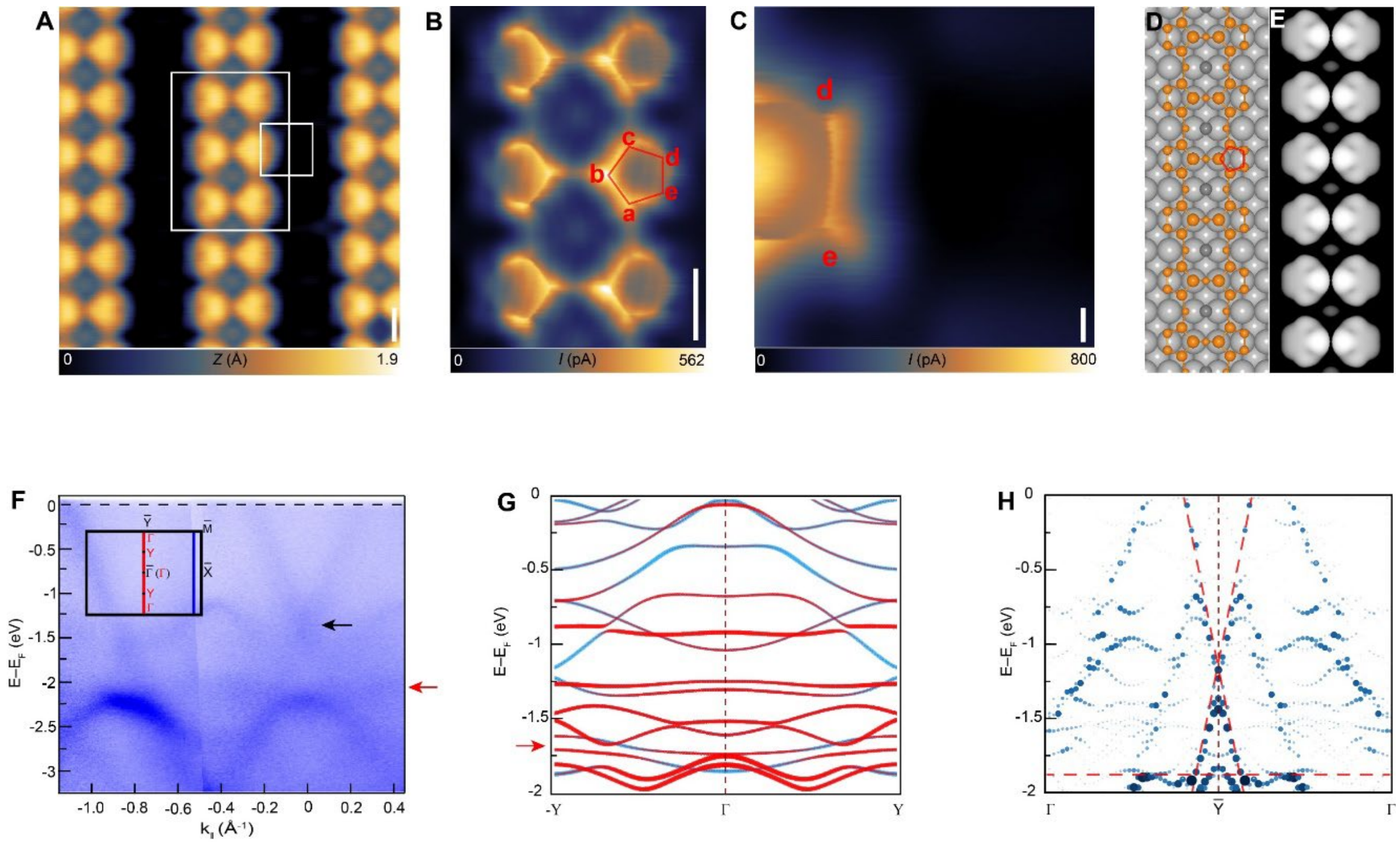

**Figure 4.** Atomic and electronic band structures of penta-dodeca-PNRs on Ag(110). A) Close-up STM images of penta-dodeca-PNRs ($V = 1$ mV, $I = 0.4$ nA). B, C) Corresponding BR-STM images of penta-dodeca-PNRs, taken at the white rectangles in panel (A) ($V = 1$ mV, $\Delta Z = -0.3$ Å; set point prior to turn off feedback, $V = 10$ mV, $I = 0.3$ nA). D) Top view of the penta-dodeca-PNRs structure on Ag(110). The orange (gray) spheres represent the P (Ag) atoms, in which more buckled-up P atoms are represented by larger orange spheres. E) Corresponding simulated STM image based on panel (D). F) ARPES spectra of penta-dodeca-PNRs on Ag(110) along the blue line. Inset: schematic drawing of the BZs of Ag(110) (black) and penta-dodeca-PNRs (red). The blue line indicates the momentum cut was taken. G) Calculated band structures of the penta-dodeca-PNRs on one layer Ag(110), showing a flat band indicated by the red arrow. The red and blue dotted lines correspond to the contributions of P and Ag atoms, respectively. H) Calculated band structures projected to PNRs and topmost Ag layer along the blue line in (F). The red dashed lines indicate the flat band and linear dispersion bands at the $\overline{Y}$ point. Scale bar: 5 Å in (A, B); 0.5 Å in (C).

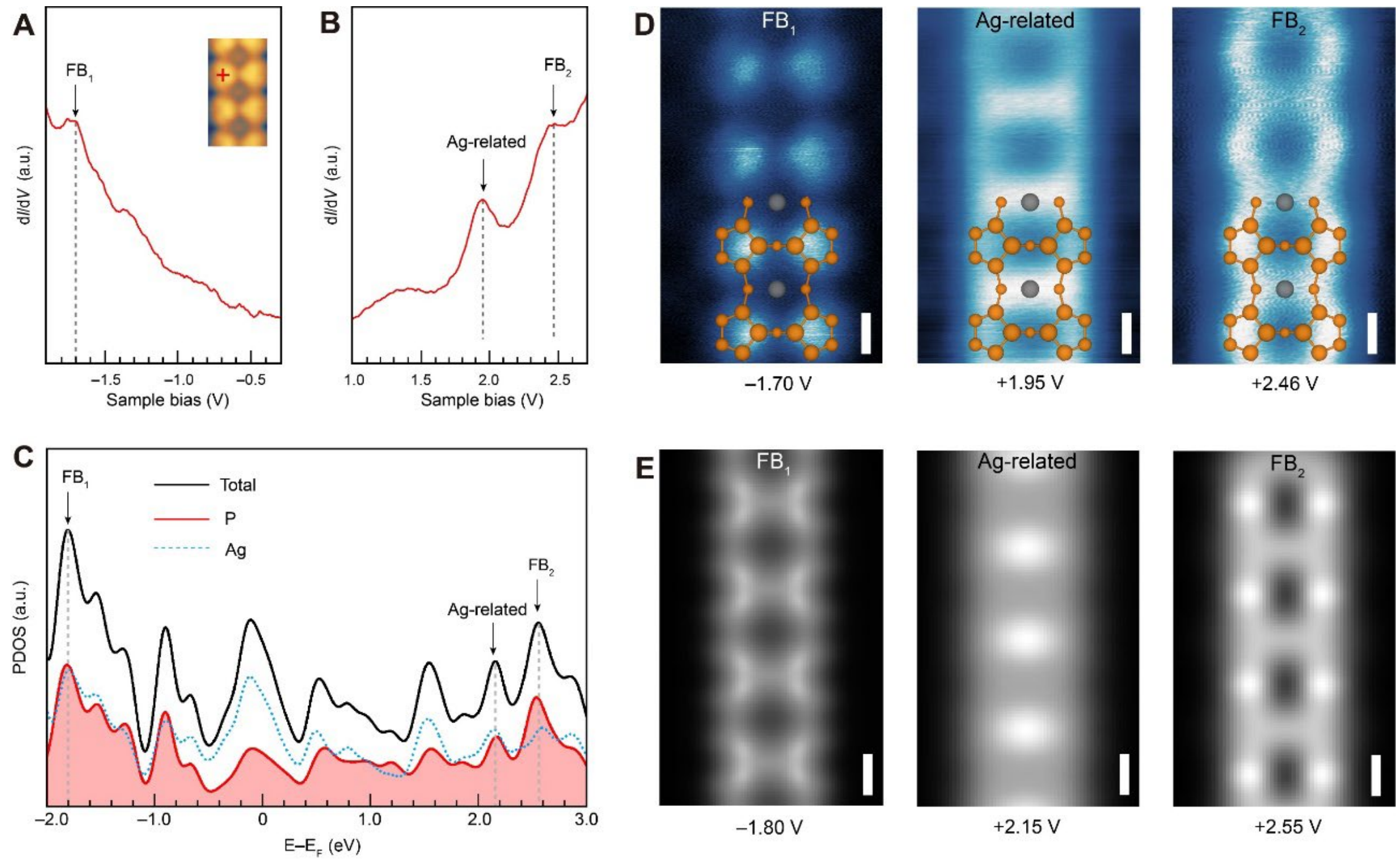

**Figure 5.** Electronic structures measurements of the penta-dodeca-PNRs on Ag(110). A, B) d*I*/d*V* point spectroscopy of the penta-dodeca-PNRs on Ag(110), taken at the red cross in the inset of (A). C) Calculated PDOS for the penta-dodeca-PNRs on one layer Ag(110). D) d*I*/d*V* maps obtained at voltages corresponding to peaks marked by arrows in (A, B), wherein the atomic structures are placed to indicate the PNRs in real space ($V = -1.70$ V, $I = 1$ nA; $V = +1.95$ V, $I = 1$ nA; $V = +2.46$ V, $I = 1$ nA). E) Calculated d*I*/d*V* maps at the energies of the peaks in (C). Scale bar: 5 Å.